**Magneto-electric operators in neutron scattering**

S W Lovesey[1, 2]

1. ISIS Facility, STFC Oxfordshire OX11 0QX, UK

2. Diamond Light Source Ltd, Oxfordshire OX11 0DE, UK

**Abstract** We succeed in deriving an exact expression for the magnetic interaction of neutrons and electrons including magneto-electric operators, allowed in the absence of a centre of inversion symmetry. Central characters are a spin anapole and an orbital (toroidal) analogue, in addition to familiar parity-even operators. A simulation of neutron diffraction by antiferromagnetic copper oxide makes full use of information inferred from a thorough investigation with resonant x-ray Bragg diffraction.

## 1. Introduction

What can be observed about a material depends on properties of the radiation employed to illuminate it. Neutrons are established as the first choice for the configuration of magnetic dipoles and distribution of magnetization, from Bragg diffraction, and spectra of magnetic excitations, from inelastic scattering.

Development of a theory of magnetic neutron scattering has been an uneven ride. Prior to the Second World War, two Nobel Laureates created two different versions of the interaction operator, denoted here by **Q**. Results by Bloch ([1], Nobel Prize 1952) and Schwinger ([2], Nobel Prize 1965) were eventually found in favour of Schwinger by Migdal [3]. (**Q** is a sum of electron spin and linear momentum field-operators and a nice derivation of it is given by Kittel [4].) Come 1953, Trammell [5] made the next big step by reporting an elegant calculation of the expectation value of **Q** suitable for the interpretation of diffraction by rare earth and actinide materials with significant amounts of orbital magnetism. A seemingly identical calculation by Johnson [6] in 1966 turned in a result that differed from Trammell's. Not long after, it was shown that Trammell's result actually contains diversionary information and the two forms are congruent [7]. In this communication, our method of working follows in the steps of Johnson's analysis [6, 8], which has been reviewed in a number of books and papers [9 - 14].

Previous calculations of matrix elements of **Q** assume equivalent unpaired electrons in an atomic shell, e.g., l = 2 (nd) or l = 3 (nf). In which case, use of a familiar dipole approximation **Q** ∝ **μ** is adequate for most purposes, where **μ** is the magnetic moment [5, 7]. In the event that symmetry of electronic states does not include a centre of spatial inversion, we prove that matrix elements include contributions from magneto-electric (acentric) operators that are both time-odd and parity-odd (angular momentum is time-odd and parity-even). The operators in question are familiar from the theory of resonant x-ray diffraction [14].

Two of the operators are anapoles, dipole operators that are essential ingredients to understanding a range of material properties [15 - 17]. A spin anapole (**s** x **n**), where **s** is the spin operator and **n** a unit vector (electric dipole), is seen to be both time-odd and parity-odd. An orbital anapole, **Ω**, with like discrete symmetries, is created by commutation of the square of total angular momentum, **L**$^2$, and (i **n**), in which the factor i = √(− 1) renders the commutator time-odd [14, 18, 19]. We find that a dipole approximation to **Q** contains operators i[**κ** x (**s** x **n**)] and i[**κ** x **Ω**], where **κ** is the direction of the scattering wavevector, in addition to the magnetic moment. Our results have application to both elastic and inelastic neutron scattering.

Consideration of the discrete symmetries of **Q** and anapoles alone shows that these constructions for it are possible, but a robust proof of the form given in this communication is necessary prior to application. Moreover, with our exact expressions the accuracy of the dipole approximation can be accessed. Matrix elements of acentric operators taken between states with angular momenta l and l' can be different from zero for l + l' odd, and they are forbidden by a centre of inversion in the symmetry operations for unpaired electrons. In the general case, the Bragg scattering amplitude is a sum of magneto-electric multipoles and familiar parity-even and time-odd multipoles. All multipoles are expressed in terms of unit tensors available from simulations of electronic structure, or derived from hand calculations.

Apology is due to the specialists on various topics. Every one of the many facets of magnetic neutron scattering changed by the new results is better known to some others than me. If, however, papers on a multi-faceted field are to be written at all, it is inevitable, since we are not immortal, that those who write such papers spend less time on any one than can be spent by a scientist who concentrates on a single facet. Some, whose scholarly austerity is unbending, will conclude that a paper covering a wide field of research should not be written at all. This communication employs a compromise, using Bragg diffraction as a means to communicate results that change the conventional interpretation of magnetic neutron scattering. Enlightened mystification as an end result is one risk with such a compromise.

The following section contains a brief summary of definitions and notation encountered in a calculation of matrix elements of the neutron-electron operator, **Q**, the subject of section 3. Additional information is relegated to an appendix. Sections 4 and 5, and the appendix, report our results for matrix elements of **Q** taken between states with orbital angular momenta, l and l'. We focus on l + l' odd, and expose the role of magneto-electric operators, although results for l + l' even will find use in some studies. A discussion of our findings, together with corresponding results for l = l', appears in section 6. A simulation of diffraction by antiferromagnetic copper oxide is reported in section 7. It makes full use of results reported by Scagnoli *et al* [17] derived from resonant x-ray Bragg diffraction. Conclusions appear in section 8.

## 2. Definitions and notation

The neutron-electron operator is a dipole, **Q**, a tensor of rank 1. Its discrete symmetries match those of a magnetic moment, because the interaction energy is the scalar product of the neutron magnetic moment and **Q**. Spherical components $Q_p$ with p = 0, ± 1 and Cartesian components are related by $Q_x = [Q_{-1} - Q_{+1}]/\sqrt{2}$, $Q_y = i[Q_{-1} + Q_{+1}]/\sqrt{2}$, and $Q_z = Q_0$.

Without loss of generality one can separate the neutron and electron parts of the interaction by using,

$$Q_p = \sum_{KK'} [\sum_{QQ'} V^K_Q(n) V^{K'}_{Q'}(e) (KQ\ K'Q' | 1p)]$$

$$= \sum_{KK'} \{V^K(n) \otimes V^{K'}(e)\}^1_p, \qquad (2.1)$$

where the second equality defines a tensor product of two operators. Here, $V^K_Q(n)$ and $V^{K'}_{Q'}(e)$ are spherical tensor operators of integer rank K and K' for neutron (n) and electronic (e) variables. It is obvious that, the ranks must obey the selection rule for a dipole, with $|K' - 1| \leq K \leq K' + 1$. Projections Q satisfy

$- K \leq Q \leq K$, a similar rule for $K'$ and $Q'$, and $(KQ\ K'Q'|1p)$ is a standard Clebsch-Gordan coefficient [20, 21]. It is often useful to work with a Wigner 3j-symbol instead of a Clebsch-Gordan coefficient, and the two are related by,

$$(a\alpha\ b\beta | IM) = (-1)^{-a+b-M} \sqrt{2I+1} \begin{pmatrix} a & b & I \\ \alpha & \beta & -M \end{pmatrix}. \quad (2.2)$$

Some relevant properties of a 3j-symbol are reported in an appendix. A matrix element of $V^{K'}_{Q'}(e)$ obeys the Wigner-Eckart Theorem [20],

$$\langle JMsl| V^{K'}_{Q'}(e) |J'M'sl'\rangle = (-1)^{J-M} (Jsl||V^{K'}(e)||J'sl') \begin{pmatrix} J & K' & J' \\ -M & Q' & M' \end{pmatrix}, \quad (2.3)$$

in which $(Jsl||V^{K'}(e)||J'sl')$ is a so-called reduced matrix-element (RME), and total angular momenta are $J = l \pm s$, $J' = l' \pm s$ with $s = 1/2$. Hereafter, we use abbreviations $\theta = Jsl$ and $\theta' = J'sl'$. If $V^{K'}(e)$ is a tensor product with electron spin and spatial variables $z^a$ and $y^b$, respectively, then,

$$(\theta||V^{K'}(e)||\theta') = (s||z^a||s)\ (l||y^b||l')\ W^{(a,b)K'}(\theta, \theta'). \quad (2.4)$$

The unit tensor $W^{(a,b)K'}(\theta, \theta')$ is available from calculations of electronic structure, and it is defined in (A.1) for a single electron [21]. The spin index $a = 0$ or 1 and $(s||z^0||s) = \sqrt{2}$ and $(s||z^1||s) = \sqrt{(3/2)}$. Our task boils down to calculating the orbital RME in (2.4) and, thereafter, transforming it to a pellucid form in which physical properties of electrons are displayed clearly.

An expectation value, or time-average, is denoted by $\langle ... \rangle$. An Hermitian operator satisfies $\langle U^{K'}_{Q'} \rangle = (-1)^{Q'} \langle U^{K'}_{-Q'} \rangle^*$ where * is the complex conjugate. The RME of a Hermitian operator obeys two fundamental identities [21],

$$(l||U^{K'}||l')^* = (-1)^{l-l'} (l'||U^{K'}||l) = \sigma_\theta \sigma_\pi (-1)^{K'} (l||U^{K'}||l'), \quad (2.5)$$

where $\sigma_\theta = \pm 1$ ($\sigma_\pi = \pm 1$) is the time-signature (parity-signature) of $U^{K'}$.

In our search for pellucid forms of $(\theta||V^{K'}(e)||\theta')$, with $l + l'$ odd, we make use of an orbital anapole created from operators for orbital angular momentum, **L**, and the electric dipole, **n**. We choose [18],

$$\mathbf{\Omega} = \mathbf{L} \times \mathbf{n} - \mathbf{n} \times \mathbf{L} = i[\mathbf{L}^2, \mathbf{n}], \quad (2.6)$$

for which $\sigma_\theta \sigma_\pi = (\sigma_\theta = -1)(\sigma_\pi = -1) = +1$. The RME,

$$(l||\Omega||l') = i\ [l(l+1) - l'(l'+1)]\ (l||n||l'), \quad (2.7)$$

is purely imaginary and symmetric with respect to an interchange of l and l', which are exact opposites of,

$$(l||n||l') = (-1)^l \sqrt{[(2l + 1)(2l' + 1)]} \begin{pmatrix} l & 1 & l' \\ 0 & 0 & 0 \end{pmatrix}, \qquad (2.8)$$

for which $\sigma_\theta \sigma_\pi = -1$. That $(l||n||l')$ vanishes unless $l + l'$ is odd is an important property. The spin anapole, $\Omega$ and **n** are Hermitian operators whose RMEs satisfy (2.5).

The RME of the angular part of the linear momentum operator $\mathbf{p} = -i\hbar\nabla$ is [20],

$$(l||p||l') = -(i\hbar/2)[2 + l'(l' + 1) - l(l + 1)](l||n||l')$$

$$= (\hbar/2)[(l||\Omega||l') - 2i(l||n||l')]. \qquad (2.9)$$

The second equality confirms that the time signature of (i **n**) matches that of $\Omega$, i.e., $\sigma_\theta = -1$ for both operators, and they can be used as operator equivalents for **p**. An orbital current and an anapole are evidently linked through (2.9).

## 3. Magnetic interaction operator

A magnetic interaction operator for neutrons, $\mathbf{Q}_\perp$, is the sum of electron spin and linear momentum field-operators [4, 11]. There is advantage in using an auxiliary operator,

$$\mathbf{Q} = \sum_j \exp(i\mathbf{R}_j \cdot \mathbf{k})[\mathbf{s}_j - (\mathbf{k} \times \nabla_j)/k^2], \qquad (3.1)$$

with $\mathbf{Q}_\perp = [\mathbf{k} \times (\mathbf{Q} \times \mathbf{k})]/k^2$. In (3.1), the sum is over all unpaired electrons in the material of interest, **k** is the scattering wavevector, and **R** and **s**, respectively, are position and spin operators, with $\nabla = i\mathbf{p}/\hbar$. We consider a single electron and subscripts j and the sum are omitted in (3.1). Use of operator equivalents (i **n**) and $\Omega$ for **p** creates dipole approximates for **Q** trailered in the Introduction.

Matrix elements for two or more equivalent electrons ($l = l'$) are best handled with fractional-parentage coefficients. In which case, the central expression (2.4) remains exactly as stated, with the unit tensor $W^{(a,b)K'}(\theta, \theta')$ replaced by a suitable generalization that incorporates the coefficients [9 - 14].

Consider expression (2.4) and the RME of the electronic tensor operator derived from (3.1). Because the second contribution to **Q** does not involve electron spin a matrix element in the Jsl basis is proportional to $W^{(0,K')K'}(\theta, \theta')$, while a matrix element of $\{\exp(i\mathbf{R}\cdot\mathbf{k})\,\mathbf{s}\}$ is proportional to $W^{(1,b)K'}(\theta, \theta')$ in the same basis.

The neutron variable in $V^K_Q(n)$ is the scattering wavevector $\mathbf{k} = k\,\kappa$. In keeping with standard practice, we actually choose to have the magnitude of the scattering vector reside in $V^K_Q(e)$, where it appears in radial integrals, and take $V^K_Q(n) = \sqrt{(4\pi)}\,Y^K_Q(\kappa)$ where $Y^K_Q(\kappa) = (-1)^Q\,\{Y^K_{-Q}(\kappa)\}^*$ is a spherical harmonic with parity signature $\sigma_\pi = (-1)^K$ [20]. In particular, $\kappa_p = \sqrt{(4\pi/3)}\,Y^1_p(\kappa)$. The rank K is odd (even) if the electronic tensor $V^{K'}_{Q'}(e)$ is parity-odd (parity-even), to preserve the parity-even status of $\mathbf{Q}$. Likewise, the electronic tensor is time-odd, because $V^K_Q(n) \propto Y^K_Q(\kappa)$ is time-even.

In Sections 4 and 5 we establish selection rules on K and K'. For example, K = K' = 1 is allowed using orbital angular momentum states l and l' with l + l' odd. The corresponding electronic dipole $V^1(e)$ is either $(\mathbf{s} \times \mathbf{n})$, $\Omega$, or $(i\,\mathbf{n})$, and $\mathbf{Q} \propto \{\kappa \otimes V^1(e)\}^1 = (i/\sqrt{2})\,[\kappa \times V^1(e)]$.

## 4. Orbital contribution

In order to calculate the orbital RME, $(l||y^{K'}||l')$ in (2.4), we start with a matrix element $\langle lm|\exp(i\mathbf{R}\cdot\mathbf{k})\,(\mathbf{k}\times\nabla)/k^2\|l'm'\rangle$ created with atomic states of the form $f_l(R)\,Y^l_m(\mathbf{n})$ with $\mathbf{R} = R\,\mathbf{n}$. (We label the radial part by the orbital angular momentum but more generally additional labels are needed.) Angular and radial operations of the gradient are separated and the corresponding RMEs are $(l||y^{K'}(a)||l')$ and $(l||y^{K'}(r)||l')$. Following initial steps deployed by Lovesey [11] in a derivation restricted to l = l', we arrive at the result,

$$(l||y^{K'}(a)||l') = (-1)^{K'+1}\,(l'||y^{K'}(a)||l)$$

$$= (i/\sqrt{2})\,\Sigma_I\,j_I(k;l,l')\,i^I\,(2K'+1)(2I+1)\sqrt{(2K+1)}\begin{pmatrix}1 & I & K\\0 & 0 & 0\end{pmatrix}\begin{Bmatrix}1 & K & K'\\I & 1 & 1\end{Bmatrix}$$

$$\times (-1)^l\,[(-1)^{l+l'+K'}\,G(I, K';l, l') - G(I, K';l', l)]. \qquad (4.1)$$

In this result, I is an integer, $G(I, K';l, l')$ is defined in (A.2) and the radial integral,

$$j_I(k;l, l') = \int_0^\infty R^2\,dR\,f_l(R)\,f_{l'}(R)\,\{j_I(kR)/kR\}, \qquad (4.2)$$

with $j_I(x)$ a spherical Bessel function. The 3j-symbol in (4.1) is different from zero for I + K odd, and the dipole selection rule K' = K, K ± 1 appears in the 6j-symbol. In addition, the result (A.2) for G(I, K';l, l') confirms $|l - l'| \leq K' \leq l + l'$. Further comments on (4.1) follow after the corresponding expression for the radial contribution to $(l||y^{K'}||l')$.

In the radial contribution the rank K' is odd for l + l' odd, and $(l||y^{K'}(r)||l')$ is purely imaginary. We find,

$(l||y^{K'}(r)||l') = (l'||y^{K'}(r)||l)$

$= - \delta_{K, K'} (1/2\sqrt{3}) i^{K'+1} g_{K'}(k;l, l') \sqrt{[K'(K' + 1)]} i (l||n||l') Z^{K'}$, (4.3)

and $(l||y^{K'}(r)||l')$ is Hermitian. In (4.3),

$$Z^{K'} = \begin{pmatrix} l & l' & K' \\ 0 & 0 & 0 \end{pmatrix} \begin{pmatrix} l & l' & 1 \\ 0 & 0 & 0 \end{pmatrix}^{-1},$$ (4.4)

and the radial integral is,

$g_{K'}(k;l, l') = - g_{K'}(k;l', l)$ (4.5)

$= (2K' + 1) \int_0^\infty R^2 \, dR \, [f_l(R) (d/dR)f_{l'}(R) - f_{l'}(R) (d/dR)f_l(R)] \{j_{K'}(kR)/Rk^2\}$.

It is evident from (4.5) that $(l||y^{K'}(r)||l')$ vanishes if radial parts of orbitals are identical, in which case $(l||y^{K'}||l') = (l||y^{K'}(a)||l')$.

The simple case l = l' is discussed at length in previous work and here we need only report the bridge to it [6, 9, 11, 14]. The radial RME is zero, while $(l||y^{K'}(a)||l)$ in (4.1) evidently vanishes unless K' is an odd integer. In previous work $(l||y^{K'}(a)||l)$ is related to A(K, K'), with K = K' ± 1 and,

A(K' − 1, K')/A(K' + 1, K') = $\sqrt{[(K' + 1)/K']}$.

Not only is K' an odd integer but properties of G(I, K'; l, l) demonstrate I = K' for l = l', which means $(l||y^{K'}(a)||l)$ is purely real and Hermitian. It can be shown that,

$\sqrt{2} (l||y^{K'}(a)||l) W^{(0,K')K'}(\theta, \theta') = (- 1)^{J'+J} \sqrt{(2J + 1)} A(K, K')$. (4.6)

Some values of A(K' − 1, K') are found in Table 1 and references [12, 13].

Hereafter, in this section, we consider a parity-odd scenario with $l + l'$ odd. Using this condition in $G(I, K';l, l')$ proves that I is even, whereupon $(l||y^{K'}(a)||l')$ like $(l||y^{K'}(r)||l')$ is purely imaginary, with $(l||y^{K'}(a)||l')$ Hermitian for K' odd. With I even it follows that K is odd and $V^K_Q(n)$ is parity-odd, as anticipated.

### 4.1 angular RME, K' even

The value K' = 0 does not occur when $l \neq l'$. We find $(l||y^{K'}(a)||l')$ is proportional to $i\,(l||n||l')$ for K' even and it is not Hermitian, unlike $(l||y^{K'}(r)||l')$ to which it is similar in other respects. It is useful to display dependence on K in the general expression derived from (A.2b); introducing,

$$t(K';l, l') = t(K';l', l)$$

$$= [(l + l' + K' + 1)\{(K')^2 - (l' - l)^2\}(l + l' - K' + 1)]^{1/2},$$

our result is,

$$(l||y^{K'}(a; K)||l') = (1/2\sqrt{3})\, i^{K'}\, j_{K'}(k;l, l')\, \sqrt{[K'/(2K' + 1)]}\, i\, (l||n||l')$$

$$\times [\{(K' + 1)/K'\}\, Z^{K' - 1}\, t(K';l, l') - Z^{K' + 1}\, t(K' + 1;l, l')], \qquad (4.7)$$

for K = K' − 1, and $(l||y^{K'}(a; K' + 1)||l') = [K'/(K' + 1)]^{1/2}\, (l||y^{K'}(a; K' − 1)||l')$. Using $l' = l − 1$, and K' = 2, K = 1 and K' = 4, K = 3,

$$(l||y^2(a; 1)||l − 1) = -\,i\, \sqrt{(5/2)}\, j_2(k;l, l')\, (l||n||l − 1)\, \sqrt{(l^2 − 1)},$$

, $\quad (l||y^4(a; 3)||l − 1) = -\,i\, (3/2)\, j_4(k;l, l')\, (l||n||l − 1) \qquad (4.8)$

$$\times [15\, (l^2 − 1)\, (l^2 − 4)/(2(4l^2 − 9))]^{1/2}.$$

### 4.2 angular RME, K' odd

The angular RME for K' odd is proportional to the orbital anapole, **Ω**, whereas the radial RME in (4.3) is proportional to (i **n**). Both RMEs vanishes unless K = K'. From (A.2b),

$$(l||y^{K'}(a)||l') = (1/2)\, \delta_{K, K'}\, i^{K' + 1}\, (l||\Omega||l')\, Z^{K'} \qquad (4.9)$$

$$\times [3K'(K' + 1)]^{-1/2}\, \{(K' + 1)\, j_{K' − 1}(k;l, l') − K'\, j_{K' + 1}(k;l, l')\}.$$

From this expression,

$$(l||y^1(a)||l') = - \delta_{K,K'} (1/2\sqrt{6}) (l||\Omega||l') \{2j_0(k;l, l') - j_2(k;l, l')\},$$

$$(l||y^3(a)||l') = \delta_{K,K'} (1/12) (l||\Omega||l') Z^3 \{4j_2(k;l, l') - 3j_4(k;l, l')\}, \quad (4.10)$$

with,

$$Z^3 = - [3(l - 1)(l + 1)/(2(2l - 3)(2l + 3))]^{1/2}, \quad (4.11)$$

for $l' = l - 1$.

For the expectation value we may write,

$$\langle \exp(i\mathbf{R}\cdot\mathbf{k})(\boldsymbol{\kappa} \times \nabla)_p \rangle = - \sqrt{(4\pi)} k \sum_{KQ} Y^K_Q(\boldsymbol{\kappa}) \sum_{K'Q'} \langle O^{K'}_{Q'} \rangle (KQ\, K'Q'|1p), \quad (4.12)$$

where a spherical tensor operator defined by,

$$(\theta||O^{K'}||\theta') = \sqrt{2} [(l||y^{K'}(a)||l') + (l||y^{K'}(r)||l')] W^{(0,K')K'}(\theta, \theta'), \quad (4.13)$$

satisfies,

$$\langle O^{K'}_{Q'} \rangle^* = (-1)^{K'+1+Q'} \langle O^{K'}_{-Q'} \rangle. \quad (4.14)$$

The unit tensor $W^{(0,K')K'}(\theta, \theta')$ is available from simulations of electronic structure, or (A.1) for a single electron. Results (4.3), (4.7) and (4.9) with $l + l'$ odd confirm explicitly that $y^{K'}(a)$ and $y^{K'}(r)$ are both parity-odd and time-odd (magneto-electric).

## 5. Orbital-spin contribution

Let us introduce an operator,

$$\Upsilon^{K'}_{Q'} = (-i)^{K+K'+1} \sqrt{(4\pi)} \{\mathbf{s} \otimes Y^K(\mathbf{n})\}^{K'}_{Q'}, \quad (5.1)$$

where the tensor product is defined in (2.1). The two operators in (5.1) commute, the phase factor makes $\Upsilon^{K'}_{Q'}$ Hermitian, and analogues of (2.5) are,

$$(\theta||\Upsilon^{K'}||\theta')^* = (-1)^{J-J'} (\theta'||\Upsilon^{K'}||\theta) = \sigma_\theta \sigma_\pi (-1)^{K'} (\theta||\Upsilon^{K'}||\theta'). \quad (5.2)$$

Note that $\Upsilon^1 = -\sqrt{(3/2)}(\mathbf{s} \times \mathbf{n})$ for $K = 1$, with $(\mathbf{s} \times \mathbf{n})$ the spin anapole. (The notation for $\Upsilon^{K'}_{Q'}$ does not make explicit its dependence on K.) From (2.4),

$$(\theta||\Upsilon^{K'}||\theta') = (-i)^{K+K'+1} \sqrt{(6\pi)} (l||Y^K(\mathbf{n})||l') W^{(1,K)K'}(\theta, \theta'). \quad (5.3)$$

This expression reveals that $(\theta||\Upsilon^{K'}||\theta')$ and $(\theta'||\Upsilon^{K'}||\theta)$ differ by the factor $(-1)^S$ with $S = 1 + l + l' + J - J' + K'$. In conjunction with identities (5.2) we find $\sigma_\theta \sigma_\pi = -(-1)^{l+l'}$ for $\Upsilon^{K'}$, which is interpreted as $\sigma_\pi = (-1)^{l+l'}$ given $\sigma_\theta = -1$.

Because $l + K + l'$ is even when $(l||Y^K(\mathbf{n})||l')$ is different from zero we have $K$ even for $l = l'$ and $K$ odd for $l + l'$ odd, and $|l - l'| \leq K \leq l + l'$. For the special case $J = J'$ and $l = l'$ the unit tensor vanishes unless $2J + K + K'$ is even, and with $2J$ odd and $K$ even one finds $K'$ odd and $K' = K \pm 1$.

The expectation value of the orbital-spin contribution to **Q**,

$$\langle \exp(i\mathbf{R}\cdot\mathbf{k})\, \mathbf{s}_p \rangle = \sqrt{(4\pi)} \sum_{KQ} Y^K_Q(\kappa) \sum_{K'Q'} i^{K'+1} \langle H^{K'}_{Q'} \rangle (KQ\, K'Q'|1p), \qquad (5.4)$$

uses a Hermitian spherical tensor operator defined by,

$$(\theta||H^{K'}||\theta') = \sqrt{[(2K'+1)/3]}\, h_K(k;l,l')\, (\theta||\Upsilon^{K'}||\theta'), \qquad (5.5)$$

whose expectation value satisfies,

$$\langle H^{K'}_{Q'} \rangle^* = (-1)^{Q'} \langle H^{K'}_{-Q'} \rangle. \qquad (5.6)$$

The radial integral in (5.5) is,

$$h_K(k;l,l') = \int_0^\infty R^2\, dR\, f_l(R)\, f_{l'}(R)\, j_K(kR). \qquad (5.7)$$

The actual significance of contributions to scattering might be determined by magnitudes of $h_K(k;l,l')$, and for each $K$ there are three allowed values of $K'$.

A dipole approximation for $l + l'$ odd follows from (5.4) on using $K = K' = 1$. One finds,

$$\langle \exp(i\mathbf{R}\cdot\mathbf{k})\, \mathbf{s} \rangle \approx i\,(3/2)\, h_1(k;l,l')\, [\boldsymbol{\kappa} \times \langle (\mathbf{s} \times \mathbf{n}) \rangle], \qquad (5.8)$$

with $l + l'$ odd. For $K = 1$ there are two values of $K'$, and (5.8) does not include the contribution $K' = 2$.

There are also interesting features when $l = l'$ for which $K$ is even [26]. In the event that also $J = J'$ the rank $K'$ is limited to odd integers. However, the situation $l = l'$ and $J \neq J'$ is realized by action of the spin-orbit coupling, for example. For this particular case, acentric features are to be found with $K'$ even, namely, $K' = K$ even. The RME $(\theta||\Upsilon^K||\theta')$ is purely imaginary and vanishes for $J = J'$, as might be anticipated. For $K' = K$,

$$(\theta||\Upsilon^K||\theta') = (-i/2)(-1)^l [(2J+1)(2J'+1)/(2K+1)]^{1/2} (Js\ J's|K\ 1). \quad (5.9)$$

The Clebsch-Gordan coefficient $(Js\ J's|K\ 1) = 0$ for $J = J'$ (K even) as anticipated, and K has a minimum value 2.

Setting Q' = 0 in (5.1) we derive the results, $\Upsilon^2_0 = \sqrt{(3/2)}\ [(\mathbf{s} \times \mathbf{n})_z\ n_z]$ for K' = K = 2, and, $\Upsilon^4_0 = (\sqrt{5}/4)\ [(\mathbf{s} \times \mathbf{n})_z\ n_z\ (7n_z^2 - 3)]$ for K' = K = 4. The two expressions for $\Upsilon^K$ are seen to be time-odd ($\sigma_\theta = -1$) and parity-even ($\sigma_\pi = +1$).

## 6. Discussion

Matrix elements of the magnetic neutron-electron interaction, **Q**, are obtained from a reduced matrix element (RME) using the Wigner-Eckart Theorem (2.3). In the communication we focus on calculating and interpreting additions to a previous theory suitable for an electronic state with one value of the orbital angular momentum, i.e., equivalent electrons with angular momentum l. Let us use this simple case to start a discussion of findings for the more general case.

If we add the restriction of a J-manifold to that of a single orbital state, so J = l − s or J = l + s, the expectation value of the neutron-electron interaction operator in (3.1) is,

$$\langle Q_p \rangle = \sqrt{(4\pi)} \sum_{K'} \sum_{QQ'} Y^{K'-1}_Q(\boldsymbol{\kappa})\ ((2K'+1)/(K'+1))$$
$$\times \langle T^{K'}_{Q'}\rangle (K'-1Q\ K'Q'|1p), \quad (6.1)$$

with K' odd. The spherical tensor operator $T^{K'}$ in (6.1) is Hermitian, parity-even and time-odd ($\sigma_\theta \sigma_\pi = -1$) and identities (5.2) apply. It is defined by,

$$(\theta||T^{K'}||\theta') = (-1)^{J'+J} (2J+1)^{1/2} [A(K'-1, K') + B(K'-1, K')], \quad (6.2)$$

with θ = Jsl = θ' in the present case of J = J' and l = l'. Quantities A(K, K') and B(K, K') in (6.2) are derived from $\{-\exp(i\mathbf{R}\cdot\mathbf{k})\ (\boldsymbol{\kappa} \times \nabla)/k\}$ and $\{\exp(i\mathbf{R}\cdot\mathbf{k})\ \mathbf{s}\}$, respectively. One has K' ≤ (2l − 1) in A(K' − 1, K') whereas K' ≤ (2l + 1) in B(K' − 1, K'). Some values of A(K' − 1, K') and B(K' − 1, K') are found in Table 1, and references [12, 13]. The contribution with K' = 1 yields a dipole approximation for **Q** [5, 7],

$$\mathbf{Q}(\text{dipole}) \approx (1/2)\ [2\langle j_0\rangle\ (g-1) + \langle j_0 + j_2\rangle\ (2-g)]\ \mathbf{J}, \quad (6.3)$$

where g is the Landé factor. Note that $\mathbf{s} \equiv (g - 1) \mathbf{J}$ and $\mathbf{l} \equiv (2 - g) \mathbf{J}$. Radial integrals $\langle j_n \rangle$ in (6.3) are derived from (5.7) using radial wave-functions $f_l(R) = f_{l'}(R)$, and $\langle j_0 \rangle = 1$ and $\langle j_n \rangle = 0$ for $n \geq 2$ with $k = 0$.

Lifting the restriction of a J-manifold changes properties of B(K, K') but not those of A(K, K'). For B(K, K) with K even ($2 \leq K \leq 2l$) is allowed when $J \neq J'$, and (6.1) is replaced by,

$$\langle Q_p \rangle = (4\pi)^{1/2} [\sum_{K'} \sum_{QQ'} Y^{K'-1}{}_Q(\kappa) ((2K' + 1)/(K' + 1)) \langle T^{K'}{}_{Q'} \rangle (K' - 1Q\,K'Q' | 1p)$$

$$+ i \sum_K \sum_{QQ'} Y^K{}_Q(\kappa) \langle T^K{}_{Q'} \rangle (KQ\,KQ' | 1p)]. \qquad (6.4)$$

The additional contribution,

$$(\theta || T^K || \theta') = i(-1)^{J'+J} (2J + 1)^{1/2} B(K, K)$$

$$= (i^K/\sqrt{3}) (2K + 1) \langle j_K \rangle (\theta || \Upsilon^K || \theta'), \qquad (6.5)$$

is encountered in Section 5, where the operator $\Upsilon^K$ is defined in (5.1). In particular, B(1, 1) is proportional to the spin anapole ($\mathbf{s} \times \mathbf{n}$) although this contribution is forbidden in neutron interaction for $l = l'$. All the foregoing expressions hold for two or more equivalent electrons, using appropriate generalizations of A(K, K') and B(K, K') and the unit tensor $W^{(a,b)K'}(\theta, \theta')$ from which they can be derived [9 - 14].

Matrix elements of **Q** using two, or more, angular momenta, with $l + l'$ odd, are related to matrix elements of magneto-electric operators, and they are forbidden by a centre of inversion symmetry. Sections 4 and 5 contain complete results for expectation values, namely, results (4.12) and (5.4) that make use of magneto-electric multipoles $\langle \mathbf{O}^{K'} \rangle$ and $\langle \mathbf{H}^{K'} \rangle$. According to (4.14) and (5.6), $\langle O^{K'}{}_{Q'} \rangle$ is Hermitian for K' odd and anti-Hermitian for K' even, while $\langle H^{K'}{}_{Q'} \rangle$ is Hermitian for all K'.

A dipole approximation analogous to (6.3) is derived from (4.12) and (5.4). Acentric dipoles are (i **n**), (**s** × **n**) and an orbital anapole $\Omega$ defined in (2.7) [18]. A dipole approximation for $l + l'$ odd is,

$$\langle \mathbf{Q} \rangle \approx i \sqrt{(3/2)} [\kappa \times \langle \mathbf{O}^1 \rangle] + i (3/2) h_1(k; l, l') [\kappa \times \langle (\mathbf{s} \times \mathbf{n}) \rangle], \qquad (6.6)$$

with,

$$i \sqrt{(3/2)} \langle \mathbf{O}^1 \rangle = - (1/\sqrt{2}) \, g_1(k; l, l') \langle \mathbf{n} \rangle$$

$$- (i/2\sqrt{2}) \{2 j_0(k; l, l') - j_2(k; l, l')\} \langle \mathbf{\Omega} \rangle. \qquad (6.7)$$

Dipole approximates are unchanged by the double vector product in the definition of the auxiliary operator through $\mathbf{Q}_\perp = [\mathbf{k} \times (\mathbf{Q} \times \mathbf{k})]/k^2 = \mathbf{Q} - \boldsymbol{\kappa}(\boldsymbol{\kappa} \cdot \mathbf{Q})$.

Radial integrals merit comment. Integrals $g_1(k; l, l')$ and $j_0(k; l, l')$ that arise in the treatment of the orbital interaction, section 4, diverge in the forward direction, likewise the dipole approximation (6.6) to which they contribute. By contrast, $h_1(k; l, l')$ that arises in the treatment of the orbital-spin interaction, section 5, vanishes in the forward direction. Calculations of standard radial integrals, denoted here by $\langle j_n \rangle$, are major undertakings; e.g., calculations for an Ir ion by Kobayashi K *et al* [22]. Thus, for our new radial integrals we are content with impressions of their behaviour derived from hydrogen radial wavefunctions, and some results appear in Figure 1. They are calculated with 3d and 3p wavefunctions that possess an overlap = $- \sqrt{5}/3$, which is used as normalization of displayed results. Results for $g_1(k; l, l')$ in Figure 1 are derived from,

$$g_1(k; 3d, 3p) = - (256/5 \, \lambda) (4 + \lambda^2)^{-3},$$

with a dimensionless wavevector $\lambda = 3 a_o k$, where $a_o$ is the Bohr radius. Similar expressions are found for $j_0(k; 3d, 3p)$ and $h_1(k; 3d, 3p)$. In actual applications ligand and metal ion orbitals will admix.

## 7. Antiferromagnetic copper oxide (CuO)

By way of illustrating foregoing results we calculate a unit-cell structure factor for neutron Bragg diffraction by antiferromagnetic CuO. The calculation is based on a construct that accounts for principal features inferred from a thorough investigation of the material carried out by Scagnoli *et al* [17] using the technique of resonant x-ray Bragg diffraction. In particular, the authors report convincing evidence for the existence of anapoles (toroidal moments), which constitute an orbital current.

Subsequent investigations of dispersion and absorption in resonant Bragg diffraction of x-rays by CuO demonstrate that some features in the diffraction data can be attributed to the anisotropic optical properties of the material [23, 24]. The intensity of a Bragg spot is reproduced in an *ab initio* calculation of dispersion and absorption but the calculated azimuthal angle-dependence (intensity as a function of rotation of the crystal about the Bragg wavevector) does not match observations [23]. These findings make a case for an independent investigation of anapoles in CuO that can be achieved with magnetic neutron diffraction, as we shall see in this section.

Copper ions occupy sites 4a in the monoclinic space-group Cc (#9 unique axis b). Below 213 K, Cu moments form a collinear antiferromagnetic motif with ordering wavevector (1/2, 0, − 1/2) and magnetic dipole moments aligned with the b-axis, as illustrated in Figure 2. Anapoles are confined to the a-c plane depicted in the lower panel of Figure 2.

To accommodate the antiferromagnetic motif we use a cell of dimension 2a x b x 2c, shown in Figure 2. Cell vectors are **a** = a (sinβ, 0, cosβ), **b** = b (0, 1, 0) and **c** = c (0, 0, 1), with oblique β and a cell volume $v_o$ = (abc) sinβ. We choose (**a***, **b**, **c**) as orthogonal crystal axes and label them (x, y, z). Reciprocal lattice vectors **a*** = 2πbc (1, 0, 0)/$v_o$, **b*** = 2πac sinβ (0, 1, 0)/$v_o$ and **c*** = 2πab (− cosβ, 0, sinβ)/$v_o$ The Bragg wavevector **k** = (h**a***, k**b***, l**c***) with integer Miller indices, and h and l odd for magnetic scattering (a = 4.693Å, b = 3.428 Å and c = 5.137 Å, and β = 99.547° [25]). The ordering wavevector = abπ (c/a + cosβ, 0, − sinβ)/$v_o$ encloses an angle ϕ with **a*** and,

$$\cos\phi = (1 + (a/c) \cos\beta)/\sqrt{[1 + (a/c)^2 + 2 (a/c) \cos\beta]} = 0.686, \quad (7.1)$$

giving ϕ = − 46.72°.

We set magnetic properties in orthogonal principal axes (ξ, η, ζ) with ξ and ζ parallel to crystal a* and b-axes, respectively. A construct for magnetic properties of a Cu ion in antiferromagnetic CuO uses a normalized wavefunction that hybridizes spin up and spin down electrons in d- and p-states (l = 2, l' = 1),

$$|\psi\rangle = N \{|s, 1/2\rangle (\alpha|l, 0\rangle + \beta|l, -2\rangle) + f |s, -1/2\rangle|l', 0\rangle\}, \quad (7.2)$$

where purely real coefficients obey $\alpha^2 + \beta^2$ = 1, the mixing parameter f is complex and $N^2$ = 1/(1 + $|f|^2$). The wavefunction possesses orbital and spin moments parallel to ζ only, as required, with $\langle L_\zeta \rangle = \langle \psi|L_\zeta|\psi \rangle = - 2 N^2 \beta^2$ and $\langle S_\zeta \rangle$ = $N^2 \{1 - |f|^2\}/2$. Results derived using (7.2) that agree with observations are [17]: Bragg intensity at the $L_2$ absorption edge is zero when coefficients obey α = β√6, and on setting f = |f|exp(− iϕ) the magneto-electric quadrupole in resonant

Bragg diffraction vanishes. Using $\beta^2 = 1/7$ in $\langle L_\zeta \rangle$ one achieves the observed saturation magnetic moment 0.65 $\mu_B$ with $|f|^2 = 0.0390$. (A full account of the properties of (7.2) relevant to resonant x-ray diffraction appears in reference [19].) There is no claim that (7.2) is a unique description of observations on CuO, for there are surely many d-orbital and p-orbital constructions with a similar host of attractive features. But there are likely very few minimal constructions - three projections m = 0 & − 2 and m' = 0 in our case - that do quiet so well on all counts.

Values of $\langle L_\zeta \rangle$ and $\langle S_\zeta \rangle$ determine the neutron-electron interaction in the dipole approximation, in the absence of contributions from acentric operators. From the latter, there is a contribution from the spin anapole. Absence of acentric quantities from the orbital interaction treated in section 4 is a consequence of orthogonal spin states in (7.2). The spin anapole is a function of $\alpha$ only, because matrix elements of **n** between $|l, -2\rangle$ and $|l', 0\rangle$ are zero. Non-zero values of $\langle (\mathbf{s} \times \mathbf{n}) \rangle = \langle \psi | (\mathbf{s} \times \mathbf{n}) | \psi \rangle$ are confined to the a-c plane with,

$$\langle (\mathbf{s} \times \mathbf{n}) \rangle_x = 2\sqrt{(1/15)}\, N^2 \alpha\, f'', \quad \langle (\mathbf{s} \times \mathbf{n}) \rangle_y = 0,$$

and $\langle (\mathbf{s} \times \mathbf{n}) \rangle_z = 2\sqrt{(1/15)}\, N^2 \alpha\, f'.$ (7.3)

where $f = f' + if''$.

Intensity of a Bragg spot is determined by the expectation value of the interaction operator. We find, $\langle \mathbf{Q} \rangle = \Pi(h, k, l)\, (\langle \mathbf{Q}_x \rangle, \langle \mathbf{Q}_y \rangle, \langle \mathbf{Q}_z \rangle)$ with,

$$\langle \mathbf{Q}_x \rangle \approx 3i\, h_1\, \kappa_y\, \langle (\mathbf{s} \times \mathbf{n}) \rangle_z, \quad \langle \mathbf{Q}_y \rangle \approx (1/2)\, N^2\, [\langle j_0 \rangle_d (1 - 2\beta^2) - \langle j_0 \rangle_p |f|^2],$$

$$\langle \mathbf{Q}_z \rangle \approx -3i\, h_1\, \kappa_y\, \langle (\mathbf{s} \times \mathbf{n}) \rangle_x, \tag{7.4}$$

and a magnetic structure factor,

$$\Pi(h, k, l) = \exp(\pi i h/4)\, [\varphi\, (\rho + 1) + \varphi^*\, (\rho - 1)\, \exp(\pi i l/2)],$$

using $\varphi = \exp(2\pi i k y)$, $y = 0.2468$, and $\rho = (-1)^k \exp(\pi i h/2)$. In $\langle \mathbf{Q}_y \rangle$ radial integrals $\langle j_0 \rangle_d$ and $\langle j_0 \rangle_p$ based on (5.7) are appropriate for the two orbital states in (7.2) and they are unity for k = 0, whereas $h_1$ in $\langle \mathbf{Q}_x \rangle$ and $\langle \mathbf{Q}_z \rangle$ vanishes in the forward direction. The anapole contribution to (7.4) contains all terms in (5.4) with K = 1, while remaining contributions to (7.4) come from the dipole approximation (6.3). Note that parity-even and parity-odd contributions differ in phase by 90°. Anapoles are absent in Bragg spots described by (7.4) and indexed (h, 0, l).

## 8. Conclusions

We provide exact matrix elements of the interaction of neutrons with unpaired electrons. Our results apply for elastic and inelastic neutron scattering, although the communication is written from a standpoint of Bragg diffraction. Evaluated for equivalent electrons, matrix elements are the same in every respect to established results [5, 6, 7, 9]. Our new matrix elements allow for different orbital angular momenta, l and l', whereas l = l' for the simpler case of equivalent electrons.

The situation l + l' odd is forbidden by a centre of inversion symmetry. In the absence of inversion, the expectation value (time-average) of the neutron-electron interaction operator for Bragg diffraction may contain contributions from electronic magneto-electric multipoles with rank K', denoted here by $\langle \mathbf{O}^{K'} \rangle$ and $\langle \mathbf{H}^{K'} \rangle$, that are parity-odd and time-odd, in addition to familiar parity-even and time-odd operators, $\langle \mathbf{T}^{K'} \rangle$ [14]. We express expectation values $\langle ... \rangle$ in terms of unit tensors available directly from simulations of electronic structure [27, 28, 29]. Magneto-electric multipoles are allowed in some centro-symmetric crystal structures. One such case is the chemical structure Pmmm suitable for some high-$T_c$ superconductors; copper ions using sites 2q in Pmmm have site symmetry mm2 that does not contain inversion and, therefore, magneto-electric multipoles are allowed.

Magneto-electric operators $\mathbf{O}^{K'}$ and $\mathbf{H}^{K'}$ are related to a spin anapole (**s** × **n**), an orbital (toroidal) counterpart, $\Omega$, and (i **n**) where **n** is the electric dipole operator and i = √(− 1). A potentially useful (dipole) approximation to the neutron-electron interaction includes operators i[κ × (**s** × **n**)] and i[κ × $\Omega$], where κ is the direction of the scattering wavevector, in addition to the magnetic moment.

We use our results to advance a case for examination of antiferromagnetic copper oxide by magnetic neutron scattering, as a means to resolve current uncertainty about anapole moments (orbital currents) in the material.

**Acknowledgements**

Entries in Table 1 were provided by Professor E Balcar, who scrutinized the first draft of the paper. Dr A N Dobrynin prepared Figure 1, and Dr V Scagnoli provided Figure 2. The presentation benefited from discussions with Dr U Staub and Professor N A Spaldin, and members of their research groups.

**Appendix**

We gather various results used or referred to in the main text. The 3j-symbol,

$$\begin{pmatrix} a & b & c \\ a & b & c \end{pmatrix},$$

vanishes unless a, b and c satisfy a triangle condition. An odd exchange of columns changes the sign by a factor $(-1)^{a+b+c}$ and,

$$\begin{pmatrix} a & b & c \\ \alpha & \beta & \gamma \end{pmatrix} = (-1)^{a+b+c} \begin{pmatrix} a & b & c \\ -\alpha & -\beta & -\gamma \end{pmatrix},$$

which means,

$$\begin{pmatrix} a & b & c \\ 0 & 0 & 0 \end{pmatrix} = 0,$$

for a + b + c odd.

The unit tensor for one electron is,

$$W^{(a,b)K'}(\theta, \theta') = [(2j+1)(2K'+1)(2j'+1)]^{1/2} \begin{Bmatrix} s & s & a \\ l & l' & b \\ j & j' & K' \end{Bmatrix}, \qquad (A.1)$$

where s = 1/2, and θ = jsl and θ' = j'sl'. The magnitude of the 9j-symbol is unchanged by an even or odd exchange of columns or rows, but an odd exchange changes its sign by a factor $(-1)^S$ with S = 1 + a + l + l' + b + j + j' + K'.

The angular orbital RME (4.1) contains,

$$G(I, K'; l, l') = \sqrt{(2l+1)} \sum_{I'} \sqrt{(2I'+1)} \, (I'||\nabla||l') \begin{pmatrix} l & I & l' \\ 0 & 0 & 0 \end{pmatrix} \begin{Bmatrix} 1 & I & K' \\ l & l' & I' \end{Bmatrix} \qquad (A.2a)$$

$$= 2(-1)^{K'} \sqrt{[6(2l+1)(2l'+1)]} \, (l'||L||l') \sum_x (2x+1)$$

$$\times \begin{pmatrix} 1 & I & x \\ 0 & 0 & 0 \end{pmatrix} \begin{Bmatrix} K' & 1 & x \\ l' & l & l' \end{Bmatrix} \begin{pmatrix} l & l' & x \\ 0 & 0 & 0 \end{pmatrix} \begin{Bmatrix} K' & 1 & x \\ 1 & I & 1 \end{Bmatrix}. \qquad (A.2b)$$

The RME (l'||∇||l') derived from (2.9) is different from zero for I' = l' ± 1, which means I is an even integer for l + l' odd. The 6j-symbol contains the triangle condition $|l - l'| \leq K' \leq l + l'$. In the second equality (l||L||l) = √[l (l + 1) (2l + 1)]. The label x is odd for l + l' odd, which makes (A.2b) particularly useful. With K' odd x has one value, x = K'. On the other hand, x = K' ± 1 for K' even, and I = K' with G(I, K';l, l') = G(I, K';l', l).

The passage between the two forms of G(I, K';l, l') in (A.2) is made by using,

$$[2 + l'(l' + 1) - I'(I' + 1)] \propto \begin{Bmatrix} I' & l' & 1 \\ 1 & 1 & l \end{Bmatrix},$$

in (l'||∇||l'). Thereafter, standard re-coupling of angular momenta leads to (A.2b) [20, 21].

Table 1. $A(K, K') = a(K, K') \langle j_{K'-1} + j_{K'+1} \rangle$, $B(K, K') = \{b(K, K')_{K'-1} \langle j_{K'-1} \rangle + b(K, K')_{K'+1} \langle j_{K'+1} \rangle\}$ with $K = K' - 1$, and $B(K, K) = b(K, K)_K \langle j_K \rangle$ used in (4.6), (6.2) and (6.5) for $(\theta||T^{K'}||\theta')$. Results are for l = 2 and l = 3, with $(\theta'||T^{K'}||\theta)$ derived from the first identity in (5.2). Radial integrals $\langle j_n \rangle$ are defined following (6.3).

$d^1$

| J = 3/2 | J' = 3/2 | J' = 5/2 |
|---|---|---|
| a(0,1) | − √(3/5) | √(1/15) |
| a(2,3) | − (4/7)√(1/5) | (2/7)√(6/5) |
| b(0, 1)$_0$ | √(1/15) | − 2√(1/15) |
| b(0, 1)$_2$ | − √(1/15) | − (1/2)√(1/15) |
| b(2, 2)$_2$ | 0 | − (5/2)√(5/42) |
| b(2, 3)$_2$ | (2/7)√(1/5) | − (8/7)√(2/15) |
| b(2, 3)$_4$ | − (12/7)√(1/5) | − (3/7)√(3/10) |
| b(4, 4)$_4$ | 0 | − (3/2)√(15/14) |

$f^1$

| J = 5/2 | J' = 5/2 | J' = 7/2 |
|---|---|---|
| a(0,1) | − (4/3)√(5/7) | (2/3)√(1/7) |
| a(2,3) | − (2/7)√(10/3) | (4/7)√(1/3) |
| b(0, 1)$_0$ | (1/3)√(5/7) | − (4/3)√(1/7) |
| b(0, 1)$_2$ | − (4/3)√(1/35) | − (1/3)√(1/7) |
| b(2, 2)$_2$ | 0 | − (1/3)√(35/6) |
| b(2, 3)$_2$ | (4/7)√(2/15) | − (16/21)√(1/3) |
| b(2, 3)$_4$ | − (2/7)√(10/3) | − (1/7)√(3) |
| b(4, 4)$_4$ | 0 | − (3/2)√(7/11) |
| b(4, 5)$_4$ | (1/11)√(10/7) | − (12/11)√(1/7) |

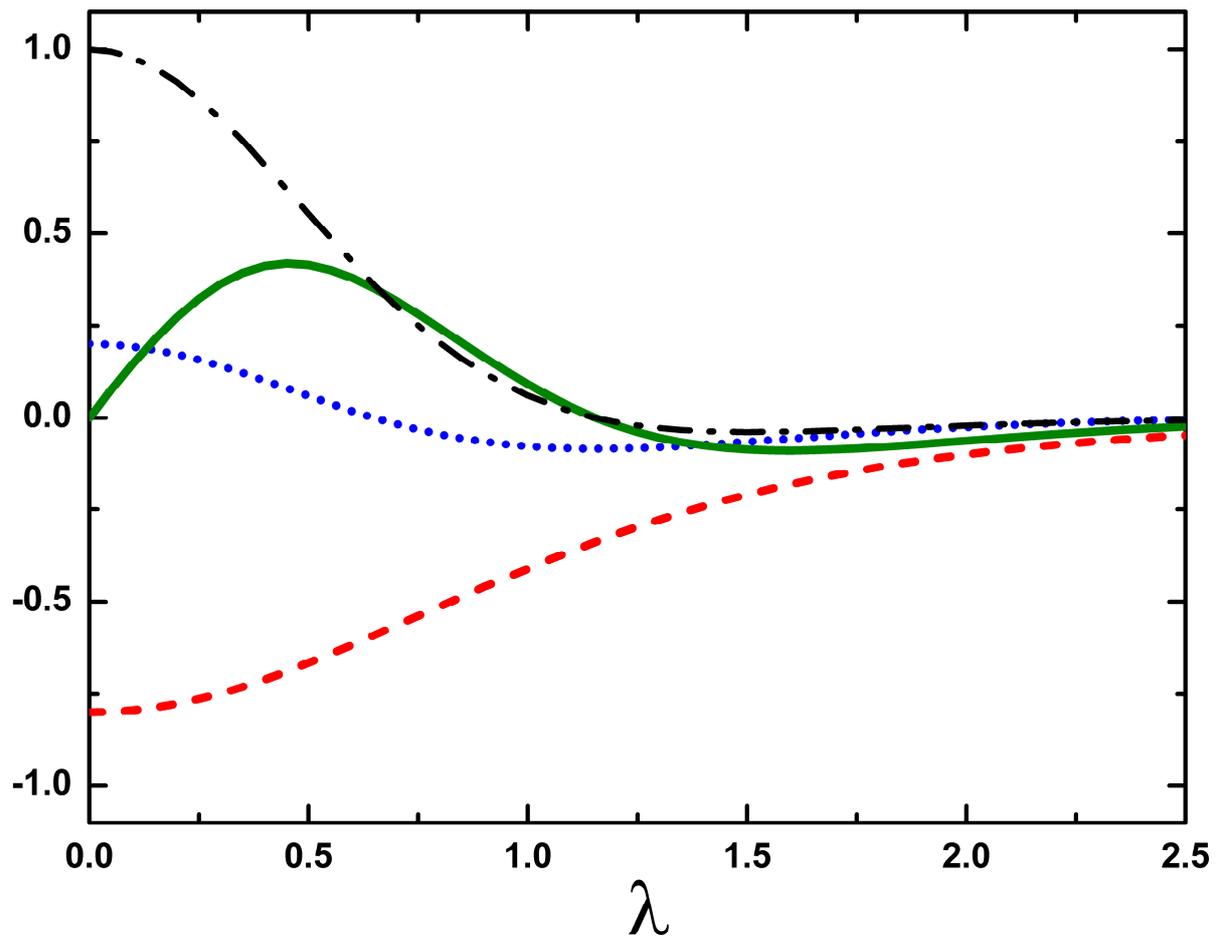

Figure 1. Radial integrals evaluated for hydrogen wavefunctions $f_{3p}(R)$ and $f_{3d}(R)$. Displayed values are normalized by the overlap of the radial functions, which is $-\sqrt{5}/3$. A dimensionless wavevector $\lambda = 3a_o k$, where $a_o$ is the Bohr radius and $k$ the magnitude of the scattering wavevector. (• • • •) $\lambda\ j_0(k; l, l')$ based on equation (4.2); (- - - -) $\lambda\ g_1(k; l, l')$ based on equation (4.5); (———) $h_1(k; l, l')$ based on equation (5.7). Also $\langle j_0 \rangle_{3d}$ is included using a dash-dot line.

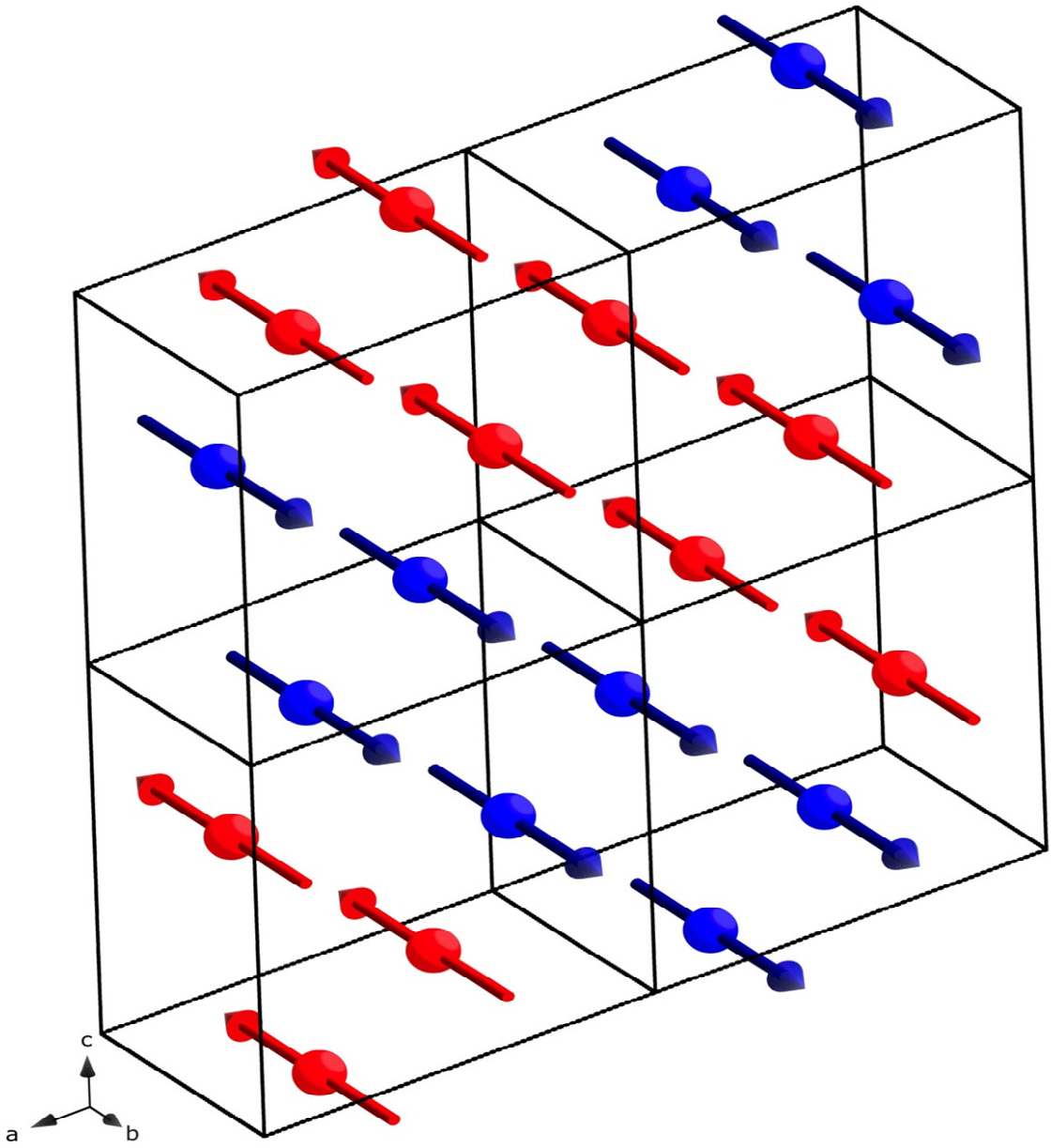

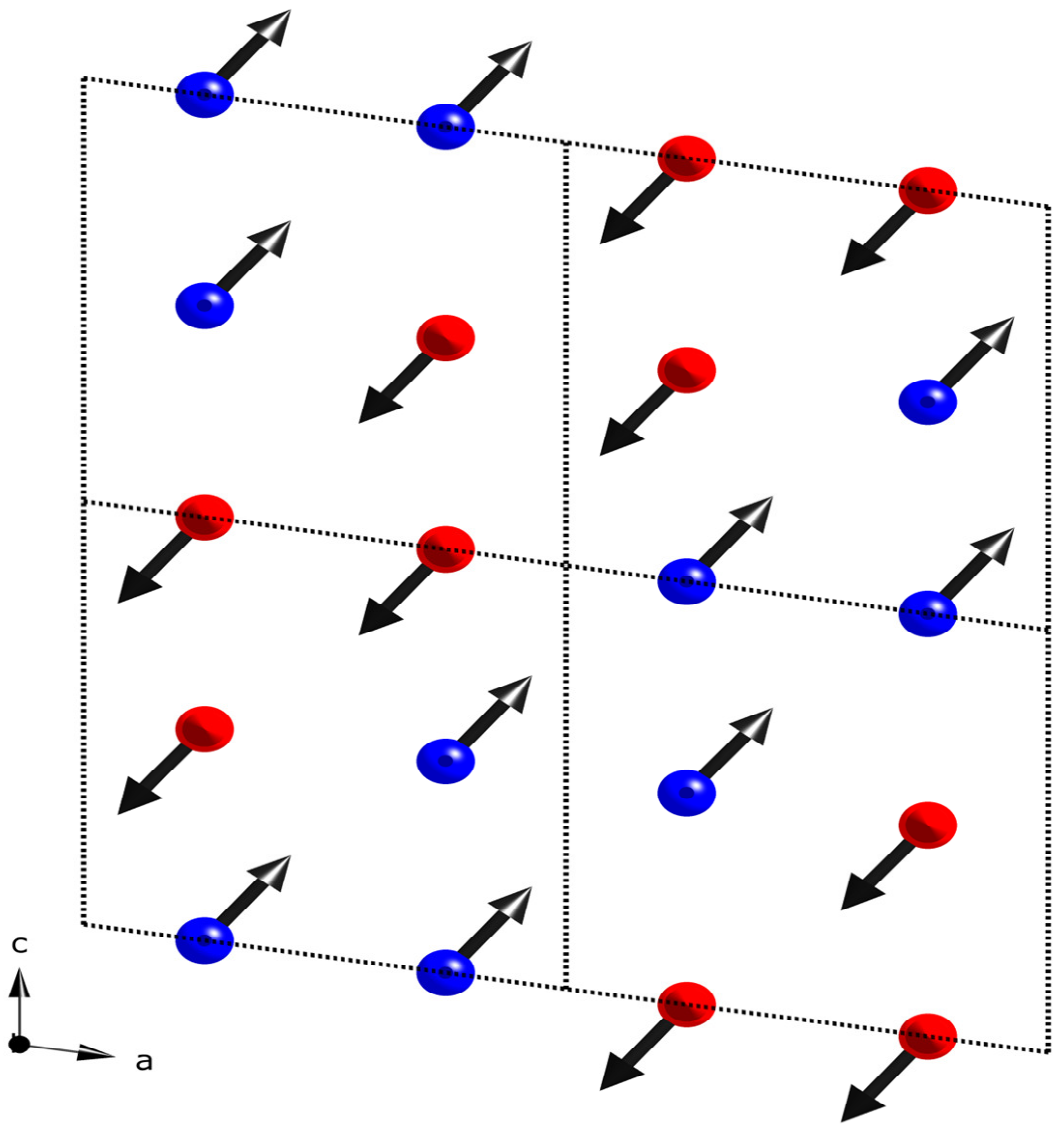

Figure 2. Upper panel depicts the magnetic unit-cell, together with the motif of magnetic dipoles in the collinear antiferromagnetic phase of CuO. Anapoles are depicted in black in the lower panel. The colour code for directions of magnetic dipoles is common to the two panels..

References

[1] Bloch F 1936 *Phys. Rev.* **50** 259 and 1937 *Phys. Rev.* **51** 944

[2] Schwinger J 1937 *Phys. Rev.* **51** 544

[3] Migdal A 1938 *Comptes Rendus Acad. Sci. URSS* **20** 551

[4] Kittel C 1987 *Quantum theory of solids* (Hoboken, NJ: Wiley)

[5] Trammell G T 1953 *Phys. Rev.* **92** 1387

[6] Johnson D F 1966 *Proc. Phys. Soc.* **88** 37

[7] Lovesey S W 1969 *J. Phys. C (Solid St. Phys.)* **2** 470

[8] Johnston D F and Rimmer D E 1969 *J. Phys. C (Solid St. Phys.)* **2** 1151

[9] Lovesey SW and Rimmer D E 1969 *Rep. Prog. Phys.* **32** 333

[10] Stassis C and Deckman H W 1975 *Phys. Rev.* B **12** 1885 and 1976 *Phys. Rev.* B **13** 4934

[11] Lovesey S W 1987 *Theory of neutron scattering from condensed matter* volume 2 (Oxford: Clarendon Press)

[12] Balcar E and Lovesey S W 1991 *J. Phys.: Condens. Matter* **3** 7095

[13] Balcar E and Lovesey S W 1989 *Theory of magnetic neutron and photon scattering* (Oxford: Clarendon Press)

[14] Lovesey S W *et al* 2005 *Phys. Rep.* **411** 233

[15] Spaldin N A *et al* 2008 *J. Phys.: Condens. Matter* **20** 434203

[16] Fernández-Rodríguez J *et al* 2010 *Phys. Rev.* **81** 085107

[17] Scagnoli V *et al* 2011 *Science* **332** 696

[18] Carra P 2004 *J. Phys. A: Math. Gen.* **37** L183

[19] Lovesey S W and Balcar E 2013 *J. Phys. Soc. Japan* **82** 021008

[20] Edmonds A R 1960 *Angular momentum in quantum mechanics* (Princeton, NJ: University Press)

[21] Balcar E and Lovesey S W 2009 *Introduction to the graphical theory of angular momentum*, Springer tracts in modern physics vol. 234 (Heidelberg: Springer)

[22] Kobayashi K *et al* 2011 *Acta Crystallogr.* A **67** 473

[23] Joly Y *et al* 2012 *Phys. Rev.* B**86** 220101(R)

[24] Lovesey S W *et al* 2014 *J. Phys.: Condens. Matter* **26** 125504

[25] Åsbrink S & Waśkowska A 1991 *J. Phys.: Condens. Matter* **3** 8173

[26] Lovesey S W and Kkalyavin D D 2014 to appear in *J. Phys.: Condens. Matter*

[27] Cricchio F *et al* 2009 *Phys. Rev. Lett.* **103** 107202

[28] Spaldin N A *et al* 2013 *Phys. Rev.* B**88** 094429

[29] Ganguly S *et al* arXiv:1312.5550v1